# Fast Switching of Bistable Magnetic Nanowires Through Collective Spin Reversal

A. Vindigni,[1] A. Rettori,[1] L. Bogani,[2] A. Caneschi,[2] D. Gatteschi,[2] R. Sessoli[2] & M.A. Novak[3]

[1]*Dipartimento di Fisica and u.d.r. INFM, Università di Firenze, I-50019 Sesto Fiorentino, Italy*

[2] *Dipartimento di Chimica and INSTM, Università di Firenze, I-50019 Sesto Fiorentino, Italy*

[3] *Instituto de Fisica Universidade Federal do Rio de Janeiro, CP68528, RJ 21945-970, Brazil*

**Abstract**

**The use of magnetic nanowires as memory units is made possible by the exponential divergence of the characteristic time for magnetization reversal at low temperature, but the slow relaxation makes the manipulation of the frozen magnetic states difficult. We suggest that finite-size segments can show a fast switching if collective reversal of the spins is taken into account. This mechanism gives rise at low temperatures to a scaling law for the dynamic susceptibility that has been experimentally observed for the dilute molecular chain Co(hfac)$_2$NitPhOMe. These results suggest a possible way of engineering nanowires for fast switching of the magnetization.**

Magnetic nanowires are nowadays key elements in nanosciences. They hold very high potential for applications in ultra-high density magnetic recording[1], logic operation devices[2], micromagnetic[3-5] and spintronics sensors[6-8]. From a fundamental point of view great interest has risen from recent realizations of molecular chain magnets[9-11] and monoatomic arrays[12]. To all these contexts the dynamics of magnetization reversal is highly relevant. For nanowires with finite, but smaller than exchange length, diameter and weak anisotropy, this phenomenon is governed by thermal nucleation and propagation of soliton-antisoliton pairs with a characteristic time that follows an Arrhenius law[13]. This behavior is observed also for genuine one-dimensional (1D) systems with high anisotropy (like the systems studied in Refs.9-12) where the reversal can be described in terms of the stochastic model proposed by Glauber in 1963[14] for the Ising chain. The consequent exponential divergence of the relaxation time at low temperatures warrants the possibility of employing nanowires as stable memory units, but, on the other side, makes their magnetic state difficult to manipulate. In this Letter, generalizing the Glauber model, we found that the dilute 1D Ising can present a faster switching if collective reversal of finite-size segments is taken into account. This mechanism gives rise to a scaling law for the dynamical susceptibility at low temperatures which we will show to be highlighted for the dilute molecular chain Co(hfac)$_2$NitPhOMe[9].

The Ising chain Hamiltonian is given by

$$H = -J\sum_{i=1}^{N-1} S_i S_{i+1} \qquad (S_i = \pm 1) \qquad (1)$$

with exchange interaction $J$, while $N$ is the number of spins. Let us consider the average time, $\tau$, the magnetization takes to be completely reversed starting from a saturated configuration. The final condition can be reached through a sequence of single spin flips, involving an activation energy $\Delta E = 4J$ for the first reversal (see Fig. 1A), and no energy cost for further flips. The relaxation time follows the Arrhenius law with $\tau \propto \exp(4J/k_BT)$. For a finite segment a different behavior is observed if $N<<\xi(T)$ where $\xi(T) \propto \exp(2J/k_BT)$ is the correlation length. In fact the energy barrier is halved, *i.e.* $\Delta E=2J$, because an edge spin is the first to be reversed (Fig. 1B)[15,16]. The complete reversal of the magnetization is attained through the propagation of the domain wall (upper part of Fig. 1C), which occurs at no energy cost.

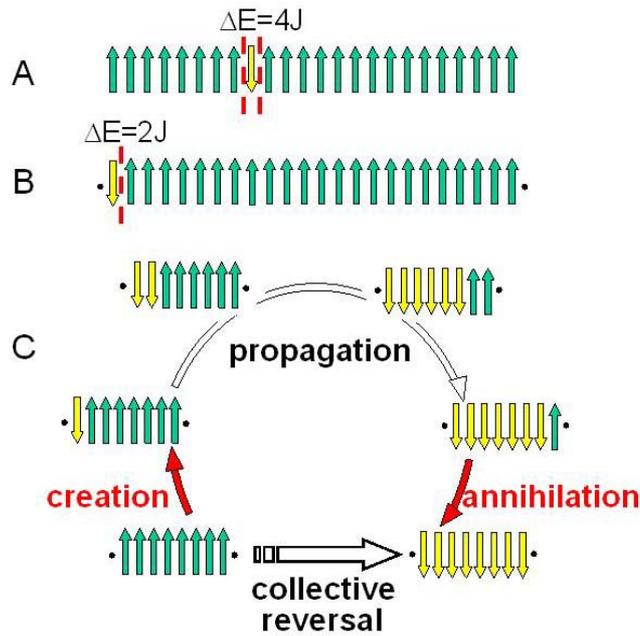

**Figure 1**: (Color online). Possible processes for the magnetization reversal. Activation energy for the reversing process starting from a bulk spin (A) or an edge spin (B). At low temperature the reversion process occurs through creation and propagation of a domain wall or by collective reversal of all the spins of a segment (C).

The probability for the domain wall to reach the other side of the segment instead of being destroyed is $\frac{1}{N-1}$[17]. Then $\tau^{-1}$ is proportional to the joint probability for the creation and propagation of a domain wall, and his prediction was experimentally confirmed for the dilute molecular chain Co(hfac)$_2$NitPhOMe[18]. Nevertheless for very high ratio $\frac{2J}{k_B T}$ (*i.e.* very low *T*) this channel is inefficient for relaxation and an alternative mechanism, the collective reversal of all spins (Fig. 1C), can become more convenient especially for short segments. In fact, if *q* is the probability to reverse a single spin in the time step $\Delta t$, the relaxation time $\tau_N$ for a segment of *N* spins is:

$$\frac{1}{\tau_N} = \frac{1}{\Delta t}\left[\frac{2q}{N-1}\exp(2J/k_B T) + q^N\right] \qquad (2)$$

We will consider both the situations of *q* obeying an Arrhenius law with its own energy barrier $\varepsilon$ and *q* being *T*-independent. Nanostructured materials usually present a narrow range of lengths[12], but often the small quantity of material does not allow a direct measurement of the magnetization dynamics. A possible alternative is the investigation of bulk crystals of 1D materials where the chains are broken by diamagnetic impurities with a random distribution giving a collection of independent segments of different lengths. Being *c* the concentration of non-magnetic impurities, the probability to have a segment of *N* spins depends on N as $(1-c)^N$. The *a.c.* susceptibility $\chi(\omega,T)$ of the doped sample is given by:

$$\chi(\omega,T) = \frac{1}{Z_c}\sum_{N=1}^{\infty}\frac{\chi_N(T)}{1-i\omega\tau_N}(1-c)^N \qquad (3)$$

where $Z_c$ is a normalizing factor, $\omega$ is the frequency, and $\chi_N(T)$ is the static susceptibility of a segment of length $N$[19]. In Fig. 2 we show the product of *T* and the real part $\chi'$ of the susceptibility defined in (3), computed for different $\omega$ with *c*= 5%. The curves change dramatically at low *T* depending on the inclusion of the collective reversal mechanism in the eq.2 for $\tau$..

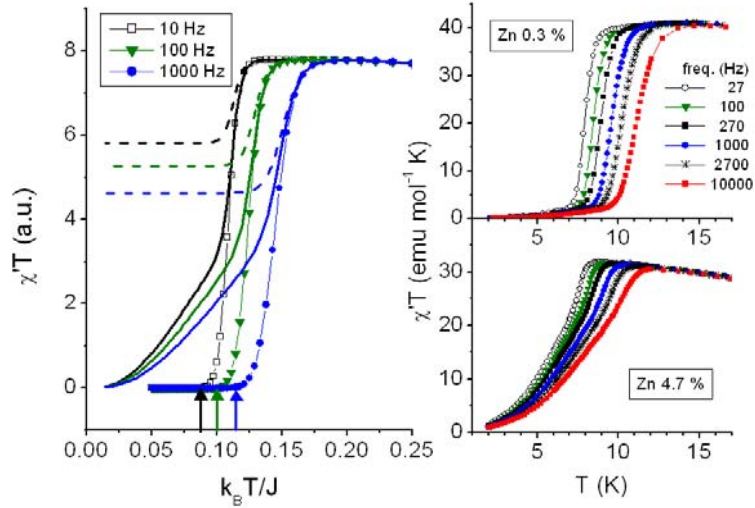

**Figure 2:** (Color online). Left: Computed product $\chi'T$ for $c = 5\%$ and 10, 100, 1000 Hz frequencies without including collective reversal in the relaxation times (lines and symbols), including collective reversal both independent of $T$ (dashed lines) and thermally activated (solid lines). Arrows correspond to the freezing temperature defined as the $k_B T/J$ ratio at which relation $\omega\tau_N = 1$ is fulfilled for $N = 2$ and collective reversal is not considered. Right: Experimental $\chi'T$ for two samples of CoPhOMe with different concentration of diamagnetic $Zn^{2+}$ doping. The *a.c.* susceptibility has been measured on single crystals in zero static field at six different frequencies.

A segment of a certain length $N$ contributes to the susceptibility only if it is able to follow the oscillating field, *i.e.* $\omega\tau_N < 1$, while for $\omega\tau_N > 1$ its contribution to the susceptibility becomes negligible. Thus the susceptibility computed without including the collective reversal channel for relaxation drops to zero below the freezing temperature, $T_f$ (Fig. 2-left). The possibility of a collective reversal with probability $q^N$ induces a non-zero susceptibility for $T < T_f$: the segments that can follow the field are those shorter than $\hat{N}$ defined by equation $q^{\hat{N}} = \omega\Delta t$. The net effect is thus to introduce a frequency dependent cut-off in eq.3. In the low $T$ limit a segment of length $N$ responds to an applied field as a whole ($N \ll \xi(T)$) and the isothermal susceptibility $\chi_N$ below

$T_f$ will be proportional to $N^2/T$. If $q$ is $T$-independent, as in the case of tunneling, $\chi'T$ tends to a plateau at low $T$ (Fig. 2-left). The height of the plateau is given by the finite sum

$$\chi'T = \frac{\mu^2}{k_B T}\frac{1}{Z_c}\sum_{N=1}^{\hat{N}} N^2(1-c)^N \qquad (4)$$

performed over all the non-frozen chains at the given $\omega$, where $\mu$ is the effective magnetic moment of the isolated center. In case of $q = \exp(-\varepsilon/k_B T)$ the cut off $\hat{N}$ will become $T$-dependent, leading to a characteristic polynomial behavior of the product $\chi'T$ for $T < T_f$. Therefore the low $T$ behavior of the product $\chi'T$ can provide a simple but powerful tool to recognize the mechanism driving the relaxation of finite Ising chains. When the collective reversal is thermally activated the $\chi'T$ behavior is characterized by a scaling law; in fact, the cut off in formula (4) is given by:

$$\hat{N} = -\log(\omega\Delta t)\frac{k_B T}{\varepsilon} \qquad (5)$$

For a given concentration of impurities $c$, the product $\chi'T$ obtained for different $\omega$ must rescale over the same curve if plotted versus the scaling variable $-\log(\omega\Delta t)T$, independent of the value of $\varepsilon$. In Fig. 2 we also report our experimental data obtained for the slow relaxing molecular Ising chain CoPhOMe, in which a non-magnetic doping agent ($Zn^{2+}$) has been inserted, with two very different concentrations, $c = 0.3\%$ and $c = 4.7\%$[15]. The low $T$ behavior of the sample with lower doping strongly reminds the calculated curves where only the Glauber dynamics is present, while that of the sample with $c = 4.7\%$ (Fig. 2-right) closely resembles the computed curves of Fig. 2 in case of thermally activated $q$. The low $T$ data of the heavily doped sample rescale over the same curve if plotted *versus* the scaling variable $-\log(\omega\Delta t)T$ (Fig. 3). This unambiguous observation of the scaling of all data over a universal curve provides, to the best of our knowledge, the first direct experimental evidence of collective reversing of a group of spins in 1D magnetic systems. This alternative channel for the relaxation of the magnetization, promoted by the presence of defects that break the chain in shorter segments, induces a significant magnetic susceptibility at $T$ well below the freezing temperature of longer chains, as clearly shown in Fig. 2.

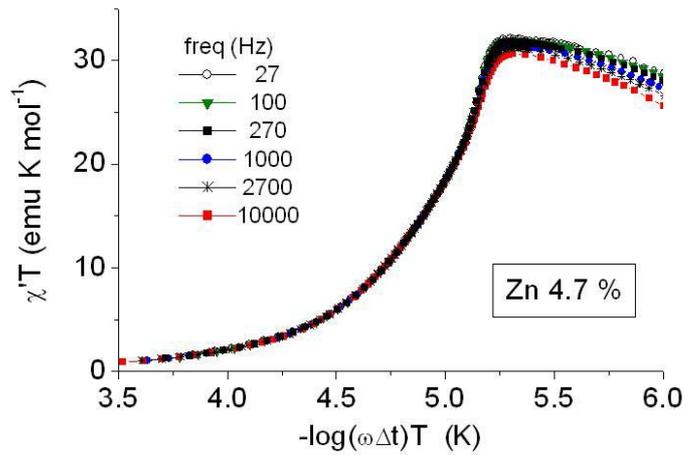

**Figure 3:** (Color online). The experimental *a.c.* susceptibility data of Fig. 2 for the more doped sample, $c$ = 4.7 %, plotted versus the scaling variable $\log(\omega \Delta t)T$ with $\Delta t$ = 5.1 ps.

The mechanism of relaxation involving collective reversal of all the spins of a short segment, here theoretically investigated and experimentally evidenced, holds great potential for fast switching of the magnetization in magnetic nanowires at low $T$, without renouncing the stability of the magnetized state. In fact it would be sufficient to randomly perturb a relatively small fraction of magnetic sites in order to break the chains in short segments that can rapidly reverse their magnetization. Removal of the perturbation would re-establish the Glauber regime, but now the clusters with reversed magnetization can eventually grow without any additional energy cost; a new frozen state with the desired magnetization can be easily reached if a weak magnetic field is applied. Several strategies to induce an instantaneous but reversible non-magnetic defect can be envisaged. For instance, blue light has been used to induce an electron transfer in Co(II)-

Fe(III) 3d cyanide-based networks, through the formation of diamagnetic Co(II)-Fe(III) low-spin pairs[20]. Magnetic nanowires are therefore very appealing for a controlled fast switching of the magnetization.

We thank the financial support from EC (RTN "QUEMOLNA"), Italian MIUR (FIRB), and German DFG.